\documentstyle[12pt,a4]{article}
\begin{document}
\renewcommand{\thefootnote}{\fnsymbol{footnote}}
\begin{titlepage}
\begin{flushright}
TUM-T31-41/93 \\
August 1993 \\
\end{flushright}
\vspace{1cm}
\begin{center}
{\Large \bf Vector and Scalar Meson Resonances in

$K\rightarrow\pi\pi\pi$ Decays} \\
\vspace{1cm}
{\large \bf S.Fajfer
\vspace{1cm}
\footnote{Alexander von Humboldt fellow}
} \\
\vspace{1cm}
{\it Physik Department, Technische Universit\"at M\"unchen\\
85748 Garching\\
FRG}
\vspace{2cm}
\end{center}
\centerline{\large \bf ABSTRACT}
\vspace{0.5cm}
  Corrections to $K\rightarrow\pi\pi\pi$ decays induced by
  vector and scalar meson exchange are investigated within chiral perturbation
theory. The widths of scalar mesons are analyzed and their influence on
$K\rightarrow\pi\pi\pi$ parameters
were examined. The overall corrections were found to be parameter dependent,
but
contributing in some cases as much as $10\%$.
\end{titlepage}

\setlength {\baselineskip}{0.75truecm}
\parindent=3pt  
\setcounter{footnote}{1}    

\newcommand{\tr}{\mbox{\rm Tr\space}}
\renewcommand{\thefootnote}{\arabic{footnote}}
\setcounter{footnote}{0}
\vspace{.5cm}
\begin{center}
{\bf 1.Introduction}\\
\end{center}
The Chiral Perturbation Theory (CHPT) offers a successful scheme for
description of strong,
weak and electromagnetic interactions at low energies
\cite{GL,EG,EGR,KMW,KM1,IP}.
The light pseudoscalar mesons play a role of Goldstone
bosons of the $SU(3)_{L} \times SU(3)_{R}$ symmetry and transition amplitudes
expanded in powers of meson momenta and masses can be calculated using
phenomenological lagrangians \cite{GL}. Unfortunately, CHPT is not
renormalizable at each order of perturbation, so one has to consider
appropriate counterterms which depend on unknown coefficients. Gasser
and Leutwyler \cite{GL} have analyzed all possible counterterms for the
strong lagrangians to the next-to leading order $O(p^{4})$ and they
have calculated their coefficients fitting the experimental amplitudes.
The coefficients of the counterterms depend on the scale $\mu$ used to
renormalize the loop graphs. The authors of refs. \cite{EG,EGR} have
investigated the role of resonances in the strong chiral lagrangian
and they have found that counterterms are saturated by resonance exchange.\\
The weak nonleptonic kaon decays were subject of interest in theoretical and
experimental particle physics for almost forty years.
The CHPT was applied to these processes
\cite{EGR,KMW,KM1,IP,EJW,FG,AIP,HC,BBE,DD}
but number of
counterterms were found to be very large \cite{KMW}. The vector-meson exchange
contribution to $K\rightarrow\pi\pi\pi$  was studied within approach of
\cite{IP,EJW}
 and it was found that they change amplitudes by only few percent.\\
The scalar mesons were involved in chiral lagrangian in order to investigate
coupling constants of the $O(p^{4})$ \cite{EG,EGR,EJW}. Their treatment
has been a persistent problem in the hadron spectroscopy \cite
{GP,GIK,J1} and therefore there are many different approaches developed in
order to clarify presently confused nature of the known $0^{++}$ mesons
\cite{GP}. Two best known scalar mesons $f_{0}(975)$ and $a_{0}(980)$
are very often treated as  $q\bar{q}q\bar{q}$ states \cite{GIK}. This
interpretation was later
reinvestigated within quark potential model as $K\bar{K}$ molecule \cite{WI}.
Recent investigation, ref.\cite{MP}, indicate that $f_{0}$ most probably is
not $K\bar{K}$ molecule, nor an amalgam of two resonances \cite{S}, but a
conventional
Breit-Wigner-like resonance.\\
Fortunately, the CHPT does not recognize the nature of these resonances, but it
gives a possibility to accommodate them as scalar octets mixed with
scalar singlet.\\
We investigate this possibility motiveted by the fact that
$f_{0} (975)$ and $K_{0}^{*}(1430)$ are effectively present in
$K\rightarrow\pi\pi\pi$ decays, while $a_{0}(980)$ affects only the
isospin-violating contribution to these amplitudes. Namely, the work
of \cite{EG,EGR,EJW} is based on the accommodation of all scalars related
CHPT couterterm parametars using only $a_{0}(980) \rightarrow \eta \pi$ decay.
Both parts of the amplitude $\Delta I = \frac{1}{2}$ and $\Delta I  =
\frac{3}{2}$
are determined assuming vacuum-insertion approximation. We confirm the result
that the $O(p^{4})$ coorections induced by vector-meson exchange effective weak
lagrangian
do not contribute to $\Delta I = \frac{1}{2}$ part of the amplitude
\cite{IP,EJW}.
We show that $\Delta I = \frac{3}{2}$ part of the amplitude coming from
corresponding
effecitive weak lagrangian is neither  affected by vector mesons. But, scalar
mesons
affect both parts of the amplitude.\\
The factorization model (or vacuum insertion approximation)\cite{BBG,PR,HC0}
which we use to
 determine the effective lagrangians which produce the  CP conserving
amplitudes, is formulated without any relations
to resonances. However, the terms of the order of $O(p^{4})$ \cite{EG,EGR,EJW}
in the strong lagrangian are being saturated by resonance contributions, what
implies that effective weak lagrangian of the same order $O(p^{4})$ can be
also influenced by their presence.\\
The outline of the work is following: in section 2 we repeat main features
of the chiral lagrangian for strong and weak interactions, containing
resonances. In the section
3 we derive and discuss the contributions to both parts of the amplitudes
$\Delta I = \frac{1}{2}$
 and $\Delta I  = \frac{3}{2}$.\\
\vspace{2cm}
\begin{center}
{\bf 2. $O(p^{4})$  effective strong and weak lagrangians}\\
\end{center}
The strong chiral lagrangian at the lowest order $O(p^{2})$ is given by
\cite{GL}
\begin{eqnarray}
{\cal L}_{s}^{2} & = & \frac{f^{2}}{4}  {\rm tr} (D_{\mu} U D^{\mu}
U^{\dag} + \chi U^{\dag} + \chi^{\dag} U)
\label {eff1}
\end{eqnarray}
where
\begin{eqnarray}
D_{\mu} U & = &\partial_{\mu} U + i ({\rm v}_{\mu} + {\rm a}_{\mu} )U + i U
({\rm v}_{\mu} -
{\rm a}_{\mu})\label{def1}
\end{eqnarray}
\begin{equation}
\chi = 2  B_{0}  ({\rm s} + i {\rm p})\label{f2}
\end{equation}
and $U = u^{2}$, is unitary $3\times 3$ matrix, with $u = exp(-\frac {i}
{\sqrt{2}}
\frac{\Phi}{f})$, $\Phi = \frac{1}{\sqrt{2}}
\sum_{i=1}^{8}\lambda_{i}\varphi^{i}$, where $\phi$ is the matrix of the
pseudoscalar fields.
The external fields ${\rm v}_{\mu}$, ${\rm a}_{\mu}$, ${\rm s}$, and ${\rm p}$
are hermitian $ 3\times 3$ matrices in
the flavor space.
The parameters $f$ and $B_{0}$ are the only free constants at
$O(p^{2})$, $f$ is
the pion constant in the chiral limit $f \simeq f_{\pi}$ and $B_{0} = <0|
\bar{u}u|0>$ . The most general lagrangian of the order $p^{4}$ is given in
the ref. \cite{EG}
\begin{eqnarray}
{\cal L}_{4} & = & l_{1} {\rm tr}(D_{\mu}U^{\dag} D^{\mu} U)^{2}
+ l_{2} {\rm tr} (D_{\mu} U^{\dag} D^{\nu}U) {\rm tr}( D_{\mu}U^{\dag}
D^{\nu}U)\nonumber\\
& + & l_{3}{\rm tr} (D_{\mu} U^{\dag} D^{\mu}U D_{\nu}U^{\dag} D^{\nu}U)
+ l_{4} {\rm tr} (D_{\mu} U^{\dag} D^{\mu}U) {\rm tr}( {\chi}^{\dag} U +{\chi}
U^{\dag})\nonumber\\
& + & l_{5}{\rm tr}( (D_{\mu} U^{\dag} D^{\mu}U) ( {\chi}^{\dag} U +{\chi}
U^{\dag}))
+l_{6}({\rm tr} ( {\chi}^{\dag} U +{\chi} U^{\dag}))^{2}\nonumber\\
& + & l_{7}({\rm tr} ( {\chi}^{\dag} U  - {\chi} U^{\dag}))^{2})
+ l_{8} ({\rm tr} ( {\chi}^{\dag} U {\chi}^{\dag} U +{\chi} U^{\dag}{\chi}
U^{\dag})\nonumber\\
& - &i l_{9} {\rm tr}(F_{R}^{\mu \nu}D_{\mu}U^{\dag} D^{\nu} U + F_{L}^{\mu
\nu}D_{\mu}U^{\dag} D^{\nu} U)
+ l_{10}(U_{R}^{\mu \nu} U F_{L \mu \nu})\nonumber\\
& + &h_{1}{\rm tr} (F_{R}^{\mu \nu} F_{R}^{\mu \nu}+ F_{L}^{\mu \nu} F_{L}^{\mu
\nu})
 +  h_{2} {\rm tr} ( {\chi}^{\dag} {\chi} )\label{i1}
\end{eqnarray}
where
\begin{equation}
F_{R,L}^{\mu \nu} = {\partial}^{\mu} (v^{\nu} \pm a^{\nu}) -
{\partial}^{\nu} (v^{\mu} \pm a^{\mu}) - i [ (v^{\mu} \pm a^{\mu}),(v^{\nu} \pm
a^{\nu})]\label{i2}
\end{equation}
$l_{1},...l_{10}$ are ten real low-energy constants which together with $f$ and
$B_{0}$ completely determine the low-energy behavior of pseudoscalar meson
interaction to
$O(p^{4})$. They arise at order $p^{4}$ and they are in general divergent
(except $l_{3}$ and $l_{7}$) \cite{GL,EG,EGR}.
They absorb the divergences of the loops arising from ${\cal L}_{2}$. It is
important to keep in mind that
they depend on a renormalization scale ${\mu}$, which does not show up in
observables.
Following the work of \cite{IP} we simply introduce vector fields in the chiral
lagrangian, even this is not the unique choice \cite{EG}:
\begin{eqnarray}
{\cal L}_{s}(V)& = &- \frac{1}{4} {\rm tr} ( {\bar V}_{\mu \nu}{\bar V}^{\mu
\nu} )
+ \frac{1}{2} M_{V}^{2}  {\rm tr} ( {\bar V}_{\mu} - \frac{i}{g}
{\Gamma}_{\mu})^{2}\label{i3}
\end{eqnarray}
Here
\begin{eqnarray}
{\bar V}_{\mu \nu} & = & {\hat V}_{\mu \nu} - i g [{\hat V}_{\mu},{\hat
V}_{\nu}]
+ \frac{i}{4 g} [ u_{\mu}, u_{\nu}] + \frac{1}{2 g} f_{\mu \nu}\label{i4}\\
\end{eqnarray}
 where ${\hat V}_{\mu \nu} = {\nabla}_{\mu} {\hat V} -{\nabla}_{\nu} {\hat
V}_{\mu}$ and
and ${\nabla}_{\mu}$ is "covariant derivative"
\begin{equation}
\nabla^{\mu} X = \partial^{\mu} X + [{\Gamma}^{\mu}, X]\label{def2}
\end{equation}
with
\begin{eqnarray}
\Gamma^{\mu} & = & \frac{1}{2} \{ u^{\dag} [\partial^{\mu} -
i({\rm v}^{\mu} + {\rm a}^{\mu})] u
+  u [\partial^{\mu} -i({\rm v}^{\mu} - {\rm a}^{\mu})] u^{\dag} \}
\label{def3}
\end{eqnarray}
 The strength ${\rm f}_{\mu \nu} = u {\rm l}_{\mu \nu} u^{\dag} + u^{\dag} {\rm
r}_{\mu \nu}$
with correspoding ${\rm l}_{\mu}$ and ${\rm r}_{\mu}$ determined as external
gauge fields of $SU(3)_{L} \times SU(3)_{R}$
as ${\rm l}_{\mu} = {\rm v}_{\mu} - {\rm a}_{\mu}$ and ${\rm r}_{\mu} = {\rm
v}_{\mu}
+ {\rm a}_{\mu}$.
Following ref. \cite{EG, EGR} from lagrangian (\ref{i3}) one dereives:
\begin{eqnarray}
{\cal L}_{s}(V)& = &- \frac{1}{4} {\rm tr} ( {\hat V}_{\mu \nu}{\hat V}^{\mu
\nu} ) +\frac{1}{2} M_{V} {\rm tr} [{\hat V}_{\mu}
{\hat V}_{\mu})\label{i5}
\end{eqnarray}
and for interacting fields
\begin{eqnarray}
{\cal L}_{int}(V) & = &  - \frac{\sqrt{2} G_{V}}{4 M_{V}} {\rm tr} ({\hat
V}_{\mu \nu}
[ u^{\mu}, u^{\nu}]) -\frac{G_{V}}{\sqrt{2} M_{V}} {\rm tr} ( f_{\mu \nu} {\hat
V}^{\mu \nu})\nonumber\\
& - & i \frac{G_{V}^{2}}{8 M_{V}^{2}} \rm tr ([ u^{\mu}, u^{\nu}] f_{\mu \nu})
+
\frac{G_{V}^{2}}{8 M_{V}^{2}} {\rm tr } ([ u_{\mu}, u_{\nu}][ u^{\mu},
u^{\nu}])\label{i6}
\end{eqnarray}
The relevant coupling constant entering into (\ref{i6}) are determined using
decay widths
$\Gamma( \rho \rightarrow e^{+} e^{-})$ and
$\Gamma( \rho \rightarrow {\pi}^{+} {\pi}^{-})$. Namely, $G_{V}$
is related to $ {\rho} \rightarrow e^{+} e^{-}$
while $F_{V}$ to ${\rho} \rightarrow {\pi}^{+} {\pi}^{-}$.
The choice of lagrangian is not unique and these  constants are
in general independent \cite{EG,EGR}, but our particular choice satisfies the
so-called the Kawarabayashi-Suzuki-Fayyazuddin-Riazuddin
relation\cite{KS,RF}
\begin{equation}
F_{V} = 2 G_{V} = \frac{M_{V}}{\sqrt{2} g}\label{k1}
\end{equation}
The kinetic term of the strong scalar lagrangian is given by
\begin{eqnarray}
{\cal L}_{k}(S) & = &\frac{1}{2} {\rm tr} (\nabla_{\mu} S \nabla^{\mu} S
- M^{2} S^{2})\label{eff2}
\end{eqnarray}
where $S$ is the scalar octet and $M_{S}$ correspondes to the scalar
masses in the chiral limit. For scalar singlet there is the kinetic term of
lagrangian
\begin{eqnarray}
{\cal L}_{k}(S_{1}) &=& \frac{1}{2} (\partial^{\mu} S_{1} \partial_{\mu}
S_{1} - M_{S_{1}}^{2} S_{1}^{2})
\label{eff3}
\end{eqnarray}
The known scalar resonances  $f_{0},a_{0}, K_{0}^{*}$ can be
described as the linear combinations of octet and singlet states. For example
$a_{0}$ and $f_{0}$ can be treated like $\rho$ and $\omega$
vector mesons
\begin{eqnarray}
f_{0} (975) & = & \frac{1}{\sqrt{3}} S_{8} +\frac{2}{\sqrt{6}} S_{1}\\
a_{0} (980) & = & -\frac{2}{\sqrt{6}} S_{1} + \frac{1}{\sqrt{3}} S_{8}
\label{def5}
\end{eqnarray}
Their interactions with Goldstone pseudoscalars can be described writing the
most general $SU(3)_{L} \times SU(3)_{R}$ lagrangian taking into
account $C$ and $P$ properties of pseudoscalars and scalars \cite{EG,EGR}
\begin{eqnarray}
{\cal L}_{int}(S) & = & c_{d}  {\rm tr} (S u_{\mu} u^{\mu}) + c_{m}  {\rm
tr}( S {\chi}_{+}) + \bar{c}_{d}  S_{1} {\rm tr} (u_{\mu} u^{\mu})
 + \bar{c}_{m}  S_{1} {\rm tr}({\chi}_{+})\label{rel1}
\end{eqnarray}
 where
\begin{eqnarray}
u_{\mu} & = & i u^{\dag} D_{\mu} U u^{\dag} \label{rel2} \\
{\chi}_{+}& = &u^{\dag} \chi u^{+} + u {\chi}^{+} u \label{rel3}
\end{eqnarray}
The experimental values of decay widths of $f_{0},a_{0}, K_{0}^{*}$ are given
in Particle Data {\bf 92}
\begin{equation}
{\Gamma}(f_{0}\rightarrow\pi\pi) = 36  {\rm MeV} \label{def6}
\end{equation}
\begin{equation}
{\Gamma}(a_{0}\rightarrow\eta\pi) = 59  {\rm MeV} \label{def7}
\end{equation}
\begin{equation}
{\Gamma}(K_{0}^{*}\rightarrow K^{-} \pi^{+}) = 267  {\rm MeV} \label{def8}
\end{equation}
Assuming the $M_{S} \simeq M_{S_{1}}$ and fitting the experimental
data for decay widths, we derive
\begin{equation}
c_{d} = 0.0220 {\rm GeV},\enspace c_{m} = 0.0288 {\rm GeV}\label{con1}
\end{equation}
and
\begin{equation}
\bar{c}_{d} = -0.0127 {\rm GeV},  \enspace \bar{c}_{m} = 0.0166 {\rm GeV}
\label{con0}
\end{equation}
These results are different then ones obtained in ref.\cite{EG} using
$a_{0}$  decay only
\begin{equation}
\mid c_{d}\mid = 0.032 {\rm GeV},\enspace \mid c_{m}\mid = 0.042 {\rm GeV}
\label{con2}
\end{equation}
\begin{equation}
c_{d} c_{m} > 0
\label{con3}
\end{equation}
 and
\begin{eqnarray}
\bar{c}_{d} & = &\frac{\epsilon}{\sqrt{3}}   c_{d},\enspace
\bar{c}_{m}  =  \frac{\epsilon}{\sqrt{3}}  c_{m} \label{con5}
\end{eqnarray}
\begin{equation}
\epsilon = \pm1\label{con6}
\end{equation}
obtained for large $N_{c}$ limit.\\
Our calculation of the coupling parametars in interacting chiral lagrangian
will lead to values of $L_{5}$ and $L_{8}$ by a factor $\sim 2$ smaller then
those in
references \cite{EG}. That will imply that these counterterms can be saturated
by contributions coming from other resonances.\\
 Following the work of Kambor et al.\cite{KMW,KM1} the CP conserving weak
lagrangian can be
written in the following form
\begin{eqnarray}
{\cal L}_{w} &= & \frac{4c_{2}}{f^{4}}  {\rm tr} ( \lambda_{6}  {\cal
J}_{\mu}  {\cal J}^{\mu}) +  \frac{4c_{3}}{f^{4}}  t_{ik}^{jl} {\rm
tr} ( Q_{j}^{i}  {\cal J}_{\mu}) {\rm tr} ( Q_{l}^{k} {\cal J}^{\mu})
\label{rl201}
\end{eqnarray}
where ${\cal J}_{\mu}$ is the weak current determined by the lowest order
expression of the left-handed current in the lagrangian (\ref{eff1}). The
couplings
$c_{2}$ and $c_{3}$ are phenomenologically determined in \cite{KM1,IP}. In
addition to vector-meson exchange analyzed in \cite{IP,EJW} there is a scalar-
meson contribution. The weak current becomes
\begin{eqnarray}
{\cal J}_{\mu} & = & u^{\dag} \{ -\frac{f^{2}}{2} u_{\mu}
 - \frac{F_{v}M_{v}}{\sqrt{2}}  V_{\mu}
 -  c_{d}  \{ u^{\mu},S \} - 2 \bar{c}_{d}  u^{\mu} S_{1} \} u
\label{eq5}
\end{eqnarray}
This expression is obtained by isolating the terms linear in $v_{\mu} -
a_{\mu}$.\\
In the further study of resonances influence on
$K\rightarrow\pi\pi\pi$ decay amplitude we use the factorized
form of the weak lagrangian. This procedure is equivalent to
evaluation of the matrix elements of four-quark weak lagrangians in
the vacuum-insertion approximation \cite{BBG}. In our case using the analysis
in ref \cite{KM1}, we take
\begin{eqnarray}
\frac{c_{2}}{f^{2}} & = & 6.6 \cdot 10^{-8}\label{con7} \\
\frac{c_{3}}{f^{2}} & = & -8.3 \cdot 10^{-10}\label{con7a}
\end{eqnarray}
It is important to point out that the coupling
constants of the chiral lagrangians are not fixed by chiral symmetry. In this
case
they are determined including next-to-leading order counterterms. The $
c_{2}/f^{2} $
is reduced by $ 30 \%$ , while $ c_{3}/f^{2} $ is unaffected by these
corrections.
The lagrangian given by factorization  approximation describes the weak
interaction of the pseudoscalars, vector-mesons and scalar-mesons.
Eliminating vector and scalar mesons by strong interaction like in \cite{EG} we
derive
 the following weak
lagrangian containing effectively vector mesons
\begin{equation}
{\cal L}_{w}^{8}(V)  = \frac{c_{2}}{M_{V}^{2}}\{ {\rm tr} (\lambda_{6} u^{\dag}
u_{\mu} u^{\nu} u_{\nu} u^{\mu} u )  -  {\rm tr} ( \lambda_{6} u^{\dag}
u_{\mu} u^{\mu} u_{\nu} u^{\nu} u ) \}\label{e19}
\end{equation}
and scalar mesons
\begin{eqnarray}
{\cal L}_{w}^{8}(S) & = & \frac{c_{2} c_{d}}{M_{S}^{2} f^{2}} 2
 \{  c_{d}  {\rm tr}( \lambda_{6} u^{\dag} \{ u_{\mu},\{ u^{\mu},u_{\nu}
u^{\nu}\}\}u)\nonumber\\
& + &c_{m} {\rm tr}(\lambda_{6} u^{\dag } \{u_{\mu} ,\{u^{\mu},\chi_{+}\}\}u)
 -  \frac{4}{3}  c_{d}  {\rm tr} (u_{\mu} u^{\mu})  {\rm tr}(\lambda_{6}
u^{\dag} u_{\mu}u^{\mu}u)\nonumber\\
& - & \frac{4}{3}  c_{m}  {\rm tr}(\chi_{+})
  {\rm tr} (\lambda_{6} u^{\dag} u_{\mu} u^{\mu} u) \}
+  \frac{\bar{c}_{d}}{M_{S}^{2} f^{2}}  2\{
\bar{c}_{d}  {\rm tr} (u_{\mu} u^{\mu})  {\rm tr}(\lambda_{6}
u^{\dag} u_{\mu}u^{\mu}u)\nonumber\\
& + &  \bar{c}_{m}  {\rm tr}(\chi_{+})
  {\rm tr} (\lambda_{6} u^{\dag} u_{\mu} u^{\mu} u) \}\label{j21}
\end{eqnarray}
The analysis of CP-invariant effective weak lagrangian transforming as
$(27_{L},1_{R})$
under $SU(3)_{L} \times SU(3)_{R}$ results in the following effective weak
lagrangians\\
\begin{eqnarray}
{\cal L}_{w}^{27}(V) & = & \frac{4 c_{3}}{f^{4}} C_{ij}^{lk}({\cal
P}_{\mu}^{3})_{ij}
( u^{\dag} u^{\mu} u)_{lk} \label{e21}
\end{eqnarray}
where
\begin{eqnarray}
{\cal P}_{\mu}^{3} & = & -{\it i}  \frac{G_{V} F_{V}}{2 M_{V}} \nabla_{\nu}
[ u^{\mu}, u_{\nu}]\label{e22}
\end{eqnarray}
for parts containing vectors, and  for part containing scalars
\begin{eqnarray}
{\cal L}_{w}^{27}(S) & = & - \{ C_{ik}^{lk} \frac{ 2 c_{d}}{M_{S}^{2}} \{
u^{\dag}
\{ u_{\mu}, j^{S}\} u -\frac{1}{3} {\rm tr} (j^{S} 2 u^{\dag} u_{\mu} u
)\}_{ik}
{\cal J}^{\mu}_{lk}\nonumber\\
& + & C_{ik}^{ln} \frac{c_{d}}{M_{S}^{2}} {\cal J}_{ik}^{\mu} \{ u^{\dag} \{
u_{\mu},j^{S}\} u - \frac{1}{3} {\rm tr}
(j^{S} 2 u^{\dag} u_{\mu} u)\}_{ln}\nonumber\\
 & + & C_{ik}^{lk}[ \frac{ 2 {\bar c}_{d}}{M_{S}^{2}} ( u^{\dag}
 u_{\mu} j^{S} u )]_{ik}{\cal J}^{\mu}_{lk}
 +  C_{ik}^{ln} \frac{ 2 {\bar c}_{d}}{M_{S}^{2}} {\cal J}_{ik}^{\mu} (
u^{\dag}
u_{\mu}j^{S} u )_{ln} \}\label{e23}
\end{eqnarray}
where $j^{S}$ is defined as
\begin{equation}
j^{S} = c_{d}  u_{\nu} u^{\nu} + c_{m}  {\chi}_{+}\label{e24}
\end{equation}
and the constants $C_{ik}^{lk}$ are determined as $C_{32}^{11} = C_{23}^{11}
=3$ and
$C_{13}^{12} = C_{31}^{21} = 1$.\\
These effective weak lagrangians are not the only source of resonance presence
in $K\rightarrow \pi \pi \pi$. There are contributions coming from
effective strong lagrangian of the $O(p^{4})$ order which counterterms are
saturated by vector and scalar-meson resonances given in the equation
(\ref{i1}).
These weak interactions occur only between pseudoscalar meson states.
In order to have complete
analysis we include in the calculation of $K\rightarrow \pi \pi \pi$ amplitude
 contributions coming from this lagrangian. The analysis of
\cite{EJW} considers the resonance contributions deriving the effective weak
lagrangians
within large-N limit approximation. Our result agrees with theirs within this
limit.\\

\vspace{1cm}
\begin{center}
{\bf 3. Effective resonance contribution to the decomposed $K\rightarrow \pi
\pi \pi$ amplitude}
\end{center}
In the notation of the reference\cite{KM1} we calculate resonance
contribution using the isospin decomposition of
$K\rightarrow\pi^{+}\pi^{0}\pi^{-}$ decay amplitude
\begin{eqnarray}
A(K\rightarrow\pi^{+}\pi^{0}\pi^{-}) & = &(\alpha_{1} +\alpha_{3})
- (\beta_{1} + \beta_{3}) Y \nonumber\\
& + &( \zeta_{1} - 2\zeta_{3}) (Y^{2} + \frac{X^{2}}{3})
+ (\xi_{1} - 2 \xi_{3}) (Y^{2} - \frac{X^{2}}{3})\label{am1}
\end{eqnarray}
with $X = \frac {1}{m_{\pi}^{2}}(s_{2} - s_{1})$ and $Y =
\frac{1}{m_{\pi}^{2}}(s_{3} - s_{0})$ where $s_{i} = (k - p_{i})^{2}$,
and $3s_{0} = s_{1}+s_{2}+s_{3}$. Here $k$ is kaon momentum, a $p_{i}$
corresponds to momentum
of $i$-th pion. The contributions coming from vector and scalar meson exchange
give
corrections to $\alpha_{1}$, $\beta_{1}$, $\zeta_{1}$, $\xi_{1}$, $\alpha_{3}$,
$\beta_{3}$, $\zeta_{3}$, $\xi_{3}$

\begin{eqnarray}
\delta \alpha_{1} & = & \frac{c_{2}}{f^{5} f_{K} M_{S}^{2}} [ m_{K}^{4}
(\frac{8}{27} c_{d}^{2} + \frac{8}{9} \bar{c}_{d}^{2} -
\frac{4}{3} c_{d} c_{m})]\label{e30}
\end{eqnarray}
\begin{eqnarray}
\delta \beta_{1} & = & - \frac{c_{2}}{f^{5} f_{K} M_{S}^{2}}  m_{\pi}^{2}
m_{K}^{2} (\frac{4}{9} c_{d}^{2} + \frac{4}{3} \bar{c}_{d}^{2} - 4
c_{d}c_{m})\nonumber\\
& - & \frac{c_{2}}{f^{2}} \frac{2 G_{V}}{M_{V}^{2} f^{3} f_{K}} m_{K}^{2}
m_{\pi}^{2}\label{e31}
\end{eqnarray}
\begin{eqnarray}
\delta \xi_{1} & = & \frac{c_{2}}{f^{5} f_{K} M_{S}^{2}}  m_{\pi}^{4}
(- \frac{2}{3} c_{d}^{2} - 2 \bar{c}_{d}^{2}) - \frac{c_{2}}{f^{2}}
\frac{2 G_{V} m_{K}^{2}}{M_{V}^{2}} \frac{m_{\pi}^{4}}{m_{K}^{2} f^{3} f_{K}}
\frac{3}{2}\label{eq32}
\end{eqnarray}
\begin{eqnarray}
\delta \zeta_{1} & = & \frac{c_{2}}{f^{5} f_{K} M_{S}^{2}}  m_{\pi}^{4}
( - \frac{2}{3} c_{d}^{2} - 2 \bar{c}_{d}^{2})\label{e33}
\end{eqnarray}
\begin{eqnarray}
\delta \alpha_{3} & = & \frac{8 c_{3}}{f^{5} f_{K} M_{S}^{2}} [ m_{K}^{4}
( - \frac{1}{18} c_{d}^{2} + \frac{5}{6} \bar{c}_{d}^{2} +
\frac{1}{3} c_{d} c_{m})]\label{e34}
\end{eqnarray}
\begin{eqnarray}
\delta \beta_{3} & = & - \frac{ 8 c_{3}}{f^{5} f_{K} M_{S}^{2}}  m_{\pi}^{2}
m_{K}^{2} (- \frac{1}{4} c_{d}^{2} + \frac{5}{4} \bar{c}_{d}^{2} - \frac{1}{4}
c_{d}c_{m})\nonumber\\
& - & \frac{ 3 c_{3}}{f^{2}} \frac{2 G_{V}}{M_{V}^{2} f^{3} f_{K}} m_{K}^{2}
m_{\pi}^{2}\label{e35}
\end{eqnarray}
\begin{eqnarray}
\delta \xi_{3} & = & - \frac{c_{3}}{2 f^{5} f_{K} M_{S}^{2}}  m_{\pi}^{4}
(- 3 c_{d}^{2} - 15 \bar{c}_{d}^{2}) - \frac{c_{3}}{f^{2}}
\frac{2 G_{V} m_{K}^{2}}{M_{V}^{2}} \frac{9 m_{\pi}^{4}}{2 m_{K}^{2} f^{3}
f_{K}}\label{eq36}
\end{eqnarray}
\begin{eqnarray}
\delta \zeta_{3} & = &- \frac{c_{3}}{2 f^{5} f_{K} M_{S}^{2}}  m_{\pi}^{4}
( - 3 c_{d}^{2} - 15 \bar{c}_{d}^{2})\label{e37}
\end{eqnarray}
The analysis of the parameters $\alpha_{1}$, $\beta_{1}$, $\xi_{1}$,
$\zeta_{1}$, $\alpha_{3}$,
$\beta_{3}$, $\xi_{3}$, and $\zeta_{3}$ was first made by Develin and Dickey
\cite{DD} and
 has been redone by Kambor et al \cite{KM1}, who have included $\Delta I =
\frac{3}{2}$ corrections to $X^{2}$
and $Y^{2}$. These two fits are basically the same. For the complete $O(p^{4})$
corrections
the loop contribution must be taken into the account.
However, the inclusion of the loops results in dependence on the
renormalization scale $\mu$. We include in
our numerical calculation results for loop contributions obtained first in ref.
\cite{KMW,KM1,EJW}. As it has been shown \cite{IP,EJW} the $K\rightarrow \pi
\pi \pi$
amplitude depends weakly on the choice of the $\mu$ scale used to renormalize
the loop graphs.\\
In our calculation we make a difference between  $f$ ( which is acctually$
\simeq f_{\pi}$) and $f_{K}$ ($f_{K} \simeq 114 {\rm MeV}$),
though this difference is not determined  by resonance counterterms
\cite{EG,EGR}.\\
The numerical results are presented in the Tables $1$ and $2$. We denote as I
the
parametars regarding scalar mesons, determined in our approach - relations
(\ref{con1}) and
(\ref{con0}), while II denotes the set of parameters determined in the paper
\cite{EG}.
We take into account both contributions : the effective resonances exchange and
the loops effect.
In the Table $1$ and $2$ we give the results for $ \mu = m_{\eta}$. The
parameters determined by complete set of scalar meson decays influence the
numerical values of isospin-decomposed $K\rightarrow\pi\pi\pi$ decay
amplitude. The $\alpha_{1}$ is still too small comparing the
experimental fit, while $\beta_{1}$ is rather unchanged by scalar
contribution. Even the overall corrections calculated using any choice
of parameters are rather small, they cannot be neglected. The precise knowledge
of
parameters describing scalar mesons is necessary in order to better understand
the role of scalar mesons in these decays, as well as in other weak,
 electromagnetic and strong processes.\\
Finally, we can comment on  CP conserving decay $K_{S} \rightarrow \pi^{+}
\pi^{0} \pi^{-}$  which is allowed even in the CP symmetry limit. This decay
rate is determined by experimentally
measured slope parameter ${\gamma}_{3}$ (see for example ref.\cite{KM1,FG})
which
in our calculation obtains the correction
\begin{eqnarray}
{\delta \gamma}_{3} & = &- \frac{4\sqrt{3}}{\sqrt{2}} \frac{c_{3}}{f^{5} f_{K}
M_{S}^{2}}
m_{\pi}^{2} m_{K}^{2} ( \frac{c_{d}^{2}}{6} + \frac{c_{d}c_{m}}{2} )\label{j50}
\end{eqnarray}
After performing the phase space integration we find
\begin{equation}
{\Gamma}(K_{S} \rightarrow \pi^{+}\pi^{0} \pi^{-})  = 1.69 \times10^{-21} {\rm
GeV}\label{j51}
\end{equation}
for the case of parameters in ref.\cite{EG}, and for parameters we derived
here, we find
\begin{equation}
{\Gamma}(K_{S} \rightarrow \pi^{+}\pi^{0} \pi^{-})  = 1.54 \times10^{-21} {\rm
GeV}\label{j52}
\end{equation}
In both cases decay width is two orders of magnitude larger than the
CP-violating contribution
arising mainly from  $K_{L} - K_{S}$ mixing, which should be measurebale in the
near future(see  e.g.\cite{FG}).\\
The conclusions of our investigations can be summarized as follows\\

(i) The corrections coming from resonance exchange are rather small, they do
depend on the choice of the parameters determined by scalar mesons data, which
still
cannot be definitely fixed due to lack of the experminetal data.\\

(ii) The analysis does not fully support phenomenological fit of \cite{KM1}
where "traces" of
vector mesons in $K\rightarrow \pi \pi \pi$ amplitude are seen,
since we found that scalar-meson corrections might be larger then vector ones.

\vspace{0.2in}
{\bf Acknowledgements}: The author thanks A.Buras  and E.De Rafael for useful
discussions. \\

\newpage
\begin{table}[h)]
\begin{center}
\begin{tabular}{|c||c|c|c|c|c|c|c|}
\hline
Amplitude    & Tree   & Loops & $I(S)$  &$II(S)$  &  $V$ &Tot$I$  &Tot$II$\\
\hline\hline
$\alpha_{1}$ & $7.80$ & $2.52$& $-0.81$ &$-0.37$  & $0$   &$9.51$ & $9.94$ \\
\hline
$\beta_{1}$  &$-1.85$ & $-1.07$& $0.24$ &$ 0.11$  &$-0.78$&$3.46$ & $3.59$ \\
\hline
$\zeta_{1}$  & $  0 $ & $-0.03$&$-0.006$&$-0.0028$& $0$   &$-0.09$&$-0.06$ \\
\hline
$\xi_{1}$    & $ 0  $ & $-0.12$&$-0.0059$&$-0.028$&$-0.09$&$-0.121$&$-0.118$ \\
\hline

$\alpha_{3}$ &$52.07$ & $-23.4$ &$2.475 $ & $1.75 $& $0$  &$31.06$ & $30.34$\\
\hline
$\beta_{3}$  &$13.9$  & $-2.28$ &$-0.22$  &$-0.10$ &$-2.3$&$11.41$ & $11.52$ \\
\hline
$\zeta_{3}$  & $  0 $ &$-0.0723$&$-0.0044$&$-0.0021$& $0$  &$-0.076$&$-0.074$
\\
\hline
$\xi_{3}$    & $ 0  $ &$-0.126$ &$0.0176$ &$0.0083$ &$-2.78$&$-0.12$&$-0.118$
\\
\hline
\end{tabular}
\end{center}
\caption{ All values are given in units $\frac{c_{2}}{f^{2}}$ for the
corrections describing
$\Delta I = \frac{1}{2}$ part of the  amplitude and in units
$\frac{c_{3}}{f^{2}}$  for the corrections coming from
$\Delta I = \frac{3}{2}$ part of the amplitude. In the first column values
coming from calculations of tree diagram, are given. In the second column there
are loop corrections calculated in ref.[5]
for the $\mu = m_{\eta}$. Third and fourth columns describe contributions of
scalar meson
resonances with fit from ref.[2] and for our fit. In the fifth column the
contribution
 determined by vector mesons
exchange is presented. The last two columns show the final corrections for two
sets of
parameters (ref.[2] and from our fit).}
\end{table}
\begin{table}[h]
\begin{center}
\begin{tabular}{|c||c|c|c|}
\hline
Amplitude    & Fit     & $ I $  & $ II$\\
\hline\hline
$\alpha_{1}$ & $91.71$ & $63.6$ & $65.6$ \\
\hline
$\beta_{1}$  & $-25.68$ & $-22.8$ & $ -23.7$\\
\hline
$\zeta_{1}$  & $  -0.047$& $ -0.059$& $-0.04$\\
\hline
$\xi_{1}$    & $ -0.151 $ & $-0.08$ & $-078$\\
\hline
$\alpha_{3}$ &$-7.36$& $-2.58$ & $-2.51$\\
\hline
$\beta_{3}$  &$2.43$ & $ 0.947$ & $9.56$ \\
\hline
$\zeta_{3}$  & $-0.021$ & $-00063$ & $ -0.0061$ \\
\hline
$\xi_{3}$    & $ -0.012$ & $-0.09$ & $ -0.098$\\
\hline
\end{tabular}
\end{center}
\caption{ All values are given in units $10^{-8}$. In the first column
we report values obtained in ref.[5] fitting the expermental data.
In the second and third column complete results for $K\rightarrow \pi \pi \pi$
amplitude
presented for  the set of parameters in ref.[2] ($I$) and for the set of
parameters derived in this paper ($II$) are given.
}
\end{table}


\newpage
\begin{thebibliography} {99}

\bibitem{GL} J.Gasser and H.Leutwyler, Nucl. Phys.{\bf B250}, 465 (1989).

\bibitem{EG} G.Ecker,J.Gasser, A.Pich and E.De Rafael, Nucl. Phys.
{\bf B321}, 311 (1989).

\bibitem{EGR} G.Ecker,J.Gasser, H.Leutwyler, A.Pich and E.De Rafael,
Phys. Lett. {\bf B223}, 425 (1990).

\bibitem{KMW} J.Kambor, J.Missimer and D.Wyler, Nucl. Phys. {\bf
B346},17 (1990).

\bibitem{KM1} J.Kambor, J.Missimer and D.Wyler, Phys. Lett. {\bf
B261}496 (1991).

\bibitem{IP} G.Isidori and A.Pugliese, Nucl. Phys. {\bf B 385}, 437 (1993).

\bibitem{EJW} G.Ecker, J.Kambor, and D.Wyler, Nucl. Phys. {\bf
B394},101 (1993).

\bibitem{GP} M.Genovese, E.Predazzi and D.B.Lichtenberg, IUHEP 248,
DFTT 14/93, April 1993.

\bibitem{GIK} F.E.Close, N.Isgur and S.Kumano, Nucl. Phys. {\bf B389}, 513
(1993).

\bibitem{J1} R.L.Jaffe, Phys. Rev. {D15} 267,281 (1977); {\em ibid}, {\bf D17}
1444 (1978).

\bibitem{WI} J.Weinstein and N.Isgur, Phys. Rev. Lett.{\bf 48} 659 (1982); {\em
ibid}, Phys. Rev. {\bf D41}, 2236 (1990).

\bibitem{BBG} W.A.Bardeen, A.J.Buras and J.M.Gerard, Phys. Lett. {\bf
B189}, 133 (1986) and references therein.

\bibitem{FG} S.Fajfer and J.-M.Gerard, Z.Phys.{\bf C42} (1989) 425.

\bibitem{AIP} G.D'Ambrosio, G.Isidori and N.Paver, Phys. Lett. {\bf B273}
(1991) 497.

\bibitem{HC} H.Y.Cheng, Phys. Rev. {\bf D42} (1990) 1579, {\bf D44} (1991) 919.

\bibitem{BBE} A.A.Belkov, G.Bohm, D.Ebert and A.Y.Lanjov, Phys.Lett. {\bf B273}
(1989) 118; Nucl. Phys. {\bf B359} (1991) 322.

\bibitem{DD} J.F.Donoghue, B.Holstein and G.Valencia, Phys.Rev.{\bf D36} (1987)
798.

\bibitem{PR} A.Pich and E.de Rafael, Nucl. Phys. {\bf B358} (1991) 496.

\bibitem{HC0} H.-Y.Cheng Phys. Rev.{\bf D42} (1990) 481.

\bibitem{KS} K.Kawarabayashi and M.Suzuki, Phys. Rev.Lett. {\bf 16} (1966) 261.

\bibitem{RF} Riazudin and Fayyazuddin, phys. Rev. {\bf 147} (1966) 255.

\bibitem{DD} T.J.Devlin and J.O.Dickey, Rev. Mod. Phys. {\bf 51} (1979) 237.

\bibitem{MP} D.Morgan and M.R.Penninton, Phys. Rev. {\bf D48} (1993) 1185.

\bibitem{S} M.Svec, A.de Lesquen and L. van Rossum, Phys. Re. {\bf D46} (1993)
949.
\end{thebibliography}
\end{document}